\documentstyle[aps,prl,psfig]{revtex}
\makeatletter
\def\setcaption#1{\def\@captype{#1}}
\makeatother

\newcommand{\SuperK}    {Super--Kamiokande}

\newcommand{\dmsq}      {$\Delta m^2$}

\begin{document}

\bigskip


\medskip
\begin{center}
\Large Tau Neutrinos Favored over Sterile Neutrinos\\
in Atmospheric Muon Neutrino Oscillations
\end{center}

{\center \it Revision of \today\\}
\bigskip
{\center \large The Super-Kamiokande Collaboration\\}
\bigskip


\begin{center}
\newcounter{foots}

\newcommand{\authoraticrr}{$^{a}$}
\newcommand{\authoratbu}{$^{b}$}
\newcommand{\authoratbnl}{$^{c}$}
\newcommand{\authoratuci}{$^{d}$}
\newcommand{\authoratcsu}{$^{e}$}
\newcommand{\authoratgmu}{$^{f}$}
\newcommand{\authoratgifu}{$^{g}$}
\newcommand{\authoratuh}{$^{h}$}
\newcommand{\authoratkek}{$^{i}$}
\newcommand{\authoratkobe}{$^{j}$}
\newcommand{\authoratkyoto}{$^{k}$}
\newcommand{\authoratlanl}{$^{l}$}
\newcommand{\authoratlsu}{$^{m}$}
\newcommand{\authoratumd}{$^{n}$}
\newcommand{\authoratsuny}{$^{o}$}
\newcommand{\authoratniigata}{$^{p}$}
\newcommand{\authoratosaka}{$^{q}$}
\newcommand{\authoratseoul}{$^{r}$}
\newcommand{\authoratshizuoka}{$^{s}$}
\newcommand{\authorattohoku}{$^{t}$}
\newcommand{\authorattokyo}{$^{u}$}
\newcommand{\authorattokai}{$^{v}$}
\newcommand{\authorattit}{$^{w}$}
\newcommand{\authoratwarsaw}{$^{x}$}
\newcommand{\authoratuw}{$^{y}$}

\newcommand{\addressoficrr}[1]{$^{a}$ #1 }
\newcommand{\addressofbu}[1]{$^{b}$ #1 }
\newcommand{\addressofbnl}[1]{$^{c}$ #1 }
\newcommand{\addressofuci}[1]{$^{d}$ #1 }
\newcommand{\addressofcsu}[1]{$^{e}$ #1 }
\newcommand{\addressofgmu}[1]{$^{f}$ #1 }
\newcommand{\addressofgifu}[1]{$^{g}$ #1 }
\newcommand{\addressofuh}[1]{$^{h}$ #1 }
\newcommand{\addressofkek}[1]{$^{i}$ #1 }
\newcommand{\addressofkobe}[1]{$^{j}$ #1 }
\newcommand{\addressofkyoto}[1]{$^{k}$ #1 }
\newcommand{\addressoflanl}[1]{$^{l}$ #1 }
\newcommand{\addressoflsu}[1]{$^{m}$ #1 }
\newcommand{\addressofumd}[1]{$^{n}$ #1 }
\newcommand{\addressofsuny}[1]{$^{o}$ #1 }
\newcommand{\addressofniigata}[1]{$^{p}$ #1 }
\newcommand{\addressofosaka}[1]{$^{q}$ #1 }
\newcommand{\addressofseoul}[1]{$^{r}$ #1 }
\newcommand{\addressofshizuoka}[1]{$^{s}$ #1 }
\newcommand{\addressoftohoku}[1]{$^{t}$ #1 }
\newcommand{\addressoftokyo}[1]{$^{u}$ #1 }
\newcommand{\addressoftokai}[1]{$^{v}$ #1 }
\newcommand{\addressoftit}[1]{$^{w}$ #1 }
\newcommand{\addressofwarsaw}[1]{$^{x}$ #1 }
\newcommand{\addressofuw}[1]{$^{y}$ #1 }

S. Fukuda\authoraticrr,
Y. Fukuda\authoraticrr,
M. Ishitsuka\authoraticrr, 
Y. Itow\authoraticrr,
T. Kajita\authoraticrr, 
J. Kameda\authoraticrr, 
K. Kaneyuki\authoraticrr,
 
K. Kobayashi\authoraticrr, 
Y. Koshio\authoraticrr, 
M. Miura\authoraticrr
% ...

R.J. Wilkes\authoratuw, 
\addtocounter{foots}{1}
K.K. Young$^{y,\fnsymbol{foots}}$


\footnotesize \it

\addressoficrr{Institute for Cosmic Ray Research, University of Tokyo, Kashiwa,
Chiba 277-8582, Japan}

\addressofbu{Department of Physics, Boston University, Boston, MA 02215, USA}

\addressofbnl{Physics Department, Brookhaven National Laboratory, 
Upton, NY 11973, USA}

\addressofuci{Department of Physics and Astronomy, University of California, 
Irvine, Irvine, CA 92697-4575, USA }

\addressofcsu{Department of Physics, California State University, 
Dominguez Hills, Carson, CA 90747, USA}

\addressofgmu{Department of Physics, George Mason University, Fairfax, VA 22030, USA }

\addressofgifu{Department of Physics, Gifu University, Gifu, Gifu 501-1193, Japan}

\addressofuh{Department of Physics and Astronomy, University of Hawaii, 
Honolulu, HI 96822, USA}

\addressofkek{Institute of Particle and Nuclear Studies, High Energy Accelerator Research Organization (KEK), Tsukuba, Ibaraki 305-0801, Japan }

\addressofkobe{Department of Physics, Kobe University, Kobe, Hyogo 657-8501, Japan}

\addressofkyoto{Department of Physics, Kyoto University, Kyoto 606-8502, Japan}

\addressoflanl{Physics Division, P-23, Los Alamos National Laboratory, 
Los Alamos, NM 87544, USA }

\addressoflsu{Department of Physics and Astronomy, Louisiana State 
University, Baton Rouge, LA 70803, USA }

\addressofumd{Department of Physics, University of Maryland, 
College Park, MD 20742, USA }

\addressofsuny{Department of Physics and Astronomy, State University of New York, Stony Brook, NY 11794-3800, USA}

\addressofniigata{Department of Physics, Niigata University, 
Niigata, Niigata 950-2181, Japan }

\addressofosaka{Department of Physics, Osaka University, 
Toyonaka, Osaka 560-0043, Japan}

\addressofseoul{Department of Physics, Seoul National University, 
Seoul 151-742, Korea}

\addressofshizuoka{Department of Systems Engineering, Shizuoka University,
Hamamatsu, Shizuoka 432-8561, Japan}

\addressoftohoku{Research Center for Neutrino Science, Tohoku University, 
Sendai, Miyagi 980-8578, Japan}

\addressoftokyo{The University of Tokyo, Tokyo 113-0033, Japan }

\addressoftokai{Department of Physics, Tokai University, Hiratsuka, 
Kanagawa 259-1292, Japan}

\addressoftit{Department of Physics, Tokyo Institute for Technology, Meguro, 
Tokyo 152-8551, Japan }

\addressofwarsaw{Institute of Experimental Physics, Warsaw University, 
00-681 Warsaw, Poland }

\addressofuw{Department of Physics, University of Washington,    
Seattle, WA 98195-1560, USA    }


\end{center}



\newpage

{\center ABSTRACT\\}

\begin{abstract}
  
  The previously published atmospheric neutrino data did not distinguish
  whether muon neutrinos were oscillating into tau neutrinos or sterile
  neutrinos, as both hypotheses fit the data.  Using data recorded in 1100
  live-days of the Super-Kamiokande detector, we use three complementary data
  samples to study the difference in zenith angle distribution due to neutral
  currents and matter effects. We find no evidence favoring sterile
  neutrinos, and reject the hypothesis at the 99\% confidence level.  On the
  other hand, we find that oscillation between muon and tau neutrinos
  suffices to explain all the results in hand.

\end{abstract}


\section{Introduction}

In previous papers\cite{evidence,SKpapers,SKupmu} Super-Kamiokande
reported evidence for the oscillation of muon neutrinos produced in
cosmic-ray induced showers in the atmosphere. This evidence rests
largely upon a strong zenith angle dependent deficit in the muon data,
which does not appear in the electron data, and hence limits the
amount of oscillation into electron neutrinos. In fact, the results
only demonstrate that muon neutrinos are disappearing, depending on
the energy and flight distance. The most plausible scenario is that
muon neutrinos oscillate to tau neutrinos, most of which are below the
3.4 GeV neutrino energy threshold for charged current tau
production. The few charged current tau events created (we
expect approximately 65 in our current sample) typically fail our cuts
that identify a single electron or muon; indeed, due to the high
energy of the interaction and the multiple decay modes of the tau, it
is difficult to isolate any set of events that can be uniquely
identified as due to charged current $\nu_\tau$ interactions.

An alternative scenario that can explain muon neutrino disappearance
is oscillation with a sterile neutrino ($\nu_s$), so named because it
has neither charged current (CC) nor neutral current (NC)
interactions.  Neutrino oscillation has also been employed to explain
two other experimental anomalies: the long-standing deficit of solar
neutrinos\cite{solarreview}, and the appearance of electron
antineutrinos in the LSND experiment\cite{LSND}. The three oscillation
signatures, LSND, atmospheric, and solar, are manifested by three
widely separated values of the mass-squared difference, $\Delta m^2 =
m^2_i - m^2_j$. Because $\Delta m^2_{31}$ must equal $\Delta m^2_{12}
+ \Delta m^2_{23}$, all three signatures cannot be accommodated with
three neutrino states. Any additional light neutrino must be sterile
to satisfy the well-known bound of three neutrino flavors that couple
to the $Z^0$\cite{z0}.

In this letter, we use more than 1000 days exposure of atmospheric
neutrino data collected by the Super-Kamiokande detector to
distinguish the behavior of $\nu_\mu \leftrightarrow \nu_\tau$
oscillation from $\nu_\mu \leftrightarrow \nu_s$ oscillation. First,
for $\nu_\mu \leftrightarrow \nu_s$ oscillation, one should observe
fewer neutral current events than for $\nu_\mu \leftrightarrow
\nu_\tau$ oscillation; in the latter case the neutral current event
rate is unchanged by oscillations.

Second, the interaction of the neutrinos with matter\cite{matter} leads to a
difference in the oscillation probability. The coherent forward
scattering of $\nu_\tau$ and $\nu_\mu$ are identical, therefore
the presence of matter in the neutrino path does not modify the
oscillation probability for a neutrino of energy $E_\nu$ that
travels a distance $L$ in vacuum:
\begin{equation}
P_{\nu_\mu \rightarrow \nu_\tau} = \sin^2 2\theta_v \sin^2 (\pi \frac{L}{l}),
\end{equation}
\noindent where $\sin^2 2\theta_v$ is the mixing angle between the
two neutrino states, and $l$ in vacuum is given by $l_v$ = $4 \pi
E_\nu / \Delta m^2$. In contrast, $\nu_s$ does not interact with
matter by definition, whereas $\nu_\mu$ does interact with matter via
the neutral current.  This introduces an effective potential which
modifies the mixing angle and oscillation length\cite{matter}:
\begin{eqnarray}
\sin^2 2\theta_{m} = \frac{\sin^2 2\theta_v}
{(\zeta - \cos 2\theta_v)^2 + \sin^2 2\theta_v},\\
l_m = \frac{l_v}{\sqrt{(\zeta - \cos 2\theta_v)^2 + \sin^2 2\theta_v}}.
\end{eqnarray}
\noindent The parameter $\zeta$ is given by $\mp \sqrt{2} E_\nu G_F
N_n / \Delta m^2$, where $N_n$ is the neutron density in the matter
traversed by the neutrino, the minus sign is for neutrinos, and the
plus sign is for antineutrinos. For the density of matter in the
earth, $\zeta$ reaches unity for $E_\nu$ of 5 GeV $\times \Delta
m^2/(10^{-3} {\rm eV^2})$. The Super-Kamiokande data indicates a
likely value for $\Delta m^2$ of $3 \times 10^{-3} {\rm eV^2}$, which
means that neutrinos with energy greater than approximately 15 GeV
will have the oscillation probability suppressed by matter effects if
the oscillation is $\nu_\mu \leftrightarrow \nu_s$. Neutrinos of lower
energy will have approximately the same oscillation probability in matter as
in vacuum, even if the oscillation is $\nu_\mu \leftrightarrow \nu_s$.

\section{The Data Sets}

Super-Kamiokande is a 50 kilotons water Cherenkov detector employing
11,146 photomultiplier tubes (PMTs) to monitor an internal detector
(ID) fiducial volume of 22.5 kilotons. Incoming and outgoing charged
particles are identified by 1885 PMTs in an optically isolated outer
volume (OD). Details of the detector, calibrations, and data reduction
can be found in Refs.\cite{evidence,SKpapers,SKupmu}.  Super-Kamiokande has
collected 9178 fully-contained (FC) events and 665 partially-contained
(PC) events in a 70.5 kiloton-year (1144 days) exposure, starting in
April 1996. FC events deposit all of their Cherenkov light in the ID
while PC events have exiting tracks which deposit some Cherenkov light
in the OD. The vertex position is reconstructed and the number of
Cherenkov rings are counted using PMT pulse height and timing
information. The directions and momenta are reconstructed and the
particle types are identified as ``$e$-like'' or ``$\mu$-like'' for
each Cherenkov ring.  In the current FC sample, there are 3107
single-ring $e$-like events, 2988 single-ring $\mu$-like events and
3083 multi-ring events.

This detector has also collected 1269 upward through-going muon (UTM)
events during 1138 live days.  Muons which leave both entrance and
exit signal clusters in the OD are regarded as through-going muon
events; those that are upward going are produced by atmospheric
neutrino interactions in the surrounding rock.  We required a minimum
track length of 7 meters in the inner detector (corresponding to a
minimum muon energy of 1.6~GeV) and a zenith angle $\cos\Theta < 0$
($\cos\Theta = -1$ means vertically upward-going events).  Because of
finite fitter resolution and multiple Coulomb scattering of muons in
the nearby rock, some down-going cosmic-ray muons appear to be coming
from $-0.1 < \cos\Theta < 0$. This background was estimated to be $9.1
\pm 0.8$ events \cite{SKupmu},
which was subtracted from the most horizontal bin.

\section{Analysis}


\subsection{Fully Contained Single-Ring Data}

First, utilizing only the FC single-ring events, we have examined the
hypotheses of two-flavor $\nu_{\mu} \leftrightarrow \nu_{\tau}$ and
$\nu_{\mu} \leftrightarrow \nu_{s}$ oscillation models using a
$\chi^{2}$ comparison of our data and Monte Carlo (MC), allowing all
important MC parameters to vary, weighted by their expected
uncertainties.  For $\nu_{\mu} \leftrightarrow \nu_{s}$, the effects
of matter on neutrino propagation through the earth were taken into
account by a numerical evolution where the density of the earth was
divided into 94 discrete steps in radius based on Ref.\cite{earth}.
Furthermore, matter effects are different with positive or negative
$\Delta m^2$ except for full mixing case. Therefore, we evaluate three
models of oscillation, (a) $\nu_\mu \leftrightarrow \nu_\tau$, (b)
$\nu_\mu \leftrightarrow \nu_s (\Delta m^{2} > 0)$, and (c) $\nu_\mu
\leftrightarrow \nu_s (\Delta m^{2} < 0)$. The data were binned by
particle type, momentum, and $\cos \Theta$. A $\chi^{2}$ is defined as:

\begin{displaymath} \chi^{2}_{FC} \equiv \sum_{\cos\Theta,p}^{65}
\left( \frac{N_{data}-N_{MC}(\sin^{2}2\theta,\Delta
m^{2},\epsilon_{j})}{\sigma} \right)^{2}+ \sum_{j} \left(
\frac{\epsilon_{j}}{\sigma_{j}} \right)^{2}, \end{displaymath}

\noindent where the sum is over five bins equally spaced in $\cos\Theta$ and
seven (six) momentum bins for $e$-like ($\mu$-like) events. $N_{data}$
is the measured number of events in each bin, $\sigma$ is the
statistical error, and $N_{MC}$ is the weighted sum of MC events. The
definition of $\chi^2$, and the treatment of systematic uncertainties,
$\epsilon_{j}$, is identical to that in Ref. \cite{evidence},
except we exclude the PC events (which we employ later in the present
report). 

The best-fit values of oscillation parameters are summarized in Table
\ref{tab:FCfits}. With the best fit parameters for $\nu_{\mu}
\leftrightarrow \nu_{\tau}$, we expect only 16 single-ring events from
$\nu_{\tau}$ charged current interactions in the current sample.
Moreover, matter induced modifications to oscillations do not produce
significant effects due to the relatively small energy ($\sim$~1~GeV)
of the parent neutrinos for the FC events. Therefore these three
hypotheses for oscillations are indistinguishable by this data sample
alone.

\nopagebreak[4]
\begin{table*}[tbp]
\caption{Best fit oscillation parameters for fully contained sample.} 
\label{tab:FCfits}
\begin{center}
\begin{tabular}{ l  c  c  c } 
Mode                        & \dmsq (eV$^2$) & $\sin^{2}2\theta$ & $\chi^{2}_{min}/d.o.f$ \\ 
\hline
$\nu_{\mu} \leftrightarrow \nu_{\tau}$                 & $3.2 \times 10^{-3}$ & 1.000 & 61.33/62 \\
$\nu_{\mu} \leftrightarrow \nu_{s} (\Delta m^{2} > 0)$ & $4.0 \times 10^{-3}$ & 0.995 & 62.56/62 \\ 
$\nu_{\mu} \leftrightarrow \nu_{s} (\Delta m^{2} < 0)$ & $3.2 \times 10^{-3}$ & 1.000 & 62.62/62 \\
\end{tabular}
\end{center}
\end{table*}
\nopagebreak[4]

\nopagebreak[4] 
\subsection{Multi-Ring Sample}

Next we employ a neutral current (NC) enriched sample of events selected
from multi-ring (MR) data. By definition a sterile neutrino does not
interact with matter even through the neutral current channel, while a
tau neutrino continues to experience the same neutral current
interactions as did the original muon neutrino.  We measure the zenith
angle distribution of NC events to distinguish between $\nu_{\mu}
\leftrightarrow \nu_{\tau}$ and $\nu_{\mu} \leftrightarrow \nu_{s}$:
if pure $\nu_{\mu} \leftrightarrow \nu_{\tau}$ oscillations are
operating, then the up/down ratio should be nearly unity; if
$\nu_{\mu} \leftrightarrow \nu_{s}$ oscillations dominate, the up/down
ratio will be measurably smaller.

In order to obtain a sample with an enhanced with NC events, we
applied the following selection criteria:
(1) vertex within the fiducial volume and no exiting track;
(2) multiple Cherenkov rings;
(3) particle identification of the brightest ring is $e$-like;
(4) visible energy greater than 400~MeV.

\begin{figure}[t]
\begin{center}
\psfig{figure=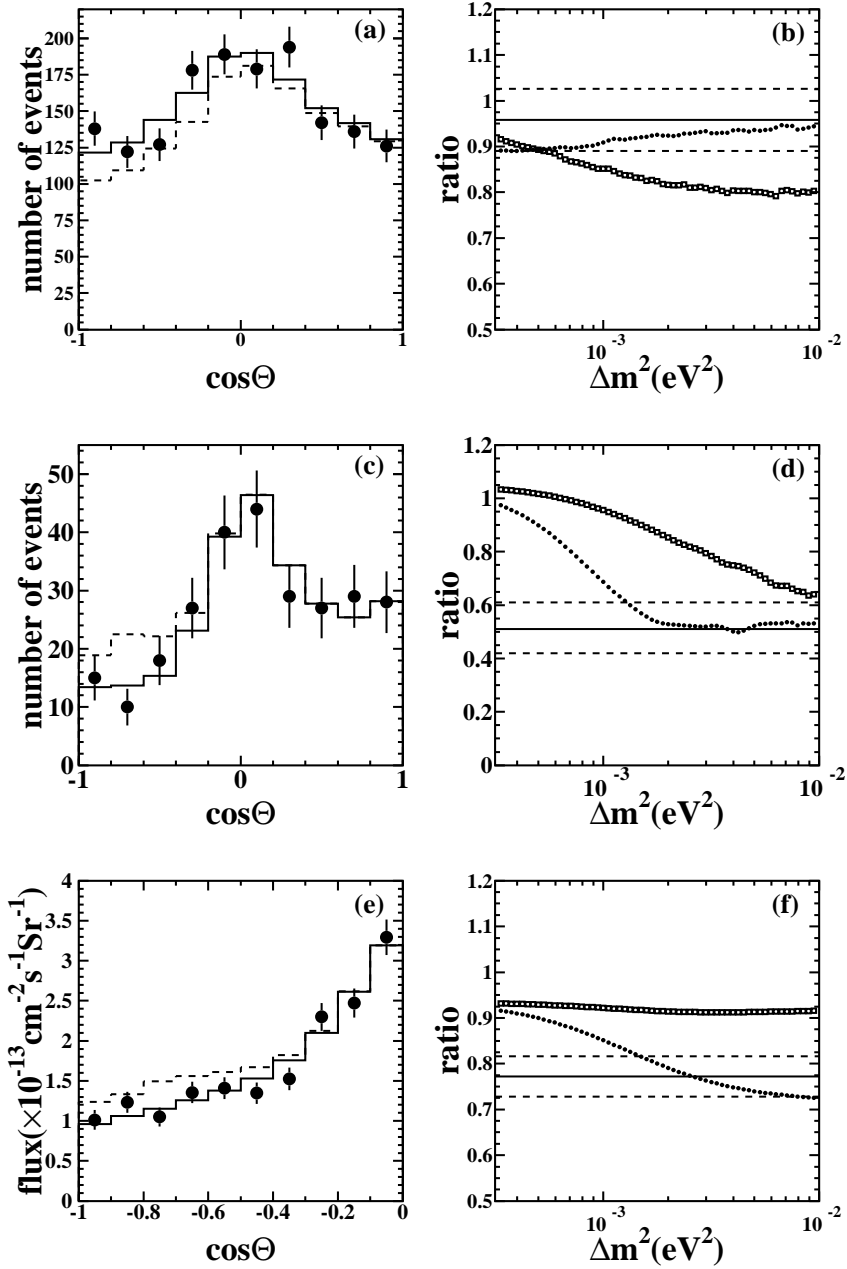,width=5.0in}
\caption{(a,c,e) Zenith angle distributions of atmospheric neutrino
events satisfying cuts described in the text: (a) multi-ring sample,
(c) partially contained sample, and (e) upward through-going muon
sample.  The black dots indicate the data and statistical errors.  The
solid line indicates the prediction for $\nu_\mu \leftrightarrow
\nu_\tau$, and the dashed for $\nu_\mu \leftrightarrow \nu_s$, with
(\dmsq,~$\sin^{2}2\theta$)=($3.2\times 10^{-3}$~eV$^{2}$,1).  The two
predictions are normalized by a common factor so that the number of
the observed events and the predicted number of events for $\nu_\mu
\leftrightarrow \nu_\tau$ are identical.  (b,d,f) Expected value of
the corresponding test ratio as a function of \dmsq. The solid
horizontal liness indicates the measured value from the Super-Kamiokande
data with statistical uncertainty indicated by dashed lines. Black
dots indicate the prediction for $\nu_\mu \leftrightarrow \nu_\tau$,
and empty squares for $\nu_\mu \leftrightarrow \nu_s$, in both
cases for maximal mixing.}
\label{fig:MR} 
\end{center}
\end{figure}

The first criterion provides a contained event sample, the second and third
criteria serve to enrich the NC event fraction.  The fourth criterion helps
to obtain good angular correlation between the incident neutrino and the
reconstructed direction, defined as the charge weighted sum of the ring
directions.  The mean angle difference between the parent neutrino and
reconstructed directions is estimated to be $33^{\circ}$.  According to our
MC study, for no oscillations (and $\nu_{\mu} \leftrightarrow \nu_{\tau}$
oscillations at best fit parameters), the resultant fraction of NC events is
29\% (30\%), $\nu_{e}$CC is 46\% (48\%) and $\nu_{\mu}$CC is 25\% (19\%) (and
$\nu_{\tau}$CC is 3\%). In contrast the FC single ring sample contains only
$\sim 6\%$ NC events. In the current exposure, 1531 events satisfy the above
criteria.  Figure \ref{fig:MR}(a) shows the zenith angle distribution of
these events with predictions from the MC.

We utilize an up-to-down ratio as the discriminant,
which cancels some systematic uncertainties (otherwise dominated by the
large uncertainty in absolute rates). 
In this context we define ``upward'' as a cosine of zenith angle less
than $-$0.4, and ``downward'' as greater than +0.4.  There are 387
upward events and 404 downward events. Figure \ref{fig:MR}(b) shows
the $\Delta m^{2}$ dependence of the expected up/down ratio in the
case of full mixing ($\sin^2 2\theta = 1)$. For $\Delta m^2$ of $3.2
\times 10^{-3}$~eV$^2$, the data are consistent with $\nu_{\mu}
\leftrightarrow \nu_{\tau}$, while the data differ from the
prediction for $\nu_{\mu} \leftrightarrow \nu_{s}$ oscillation by 2.4
standard deviations.


We estimated the total uncertainty in the up/down ratio of the data
and MC to be $\pm2.9\%$, dominated by the 2.6\% uncertainty in the
neutrino flux caused by the absorption of muons in the mountain above
the detector.  All other sources of systematic uncertainty such as
background contamination, the up/down response of the detector, the
uncertainty of the NC cross sections, and the difference between the
up/down ratio of two independent flux calculations\cite{Honda,Bartol}
are less than 1\%.

\subsection{Partially Contained Sample}

Next we report on the search for matter effects by using high energy
partially contained events. As discussed above, matter effects impact
only $\nu_{\mu} \leftrightarrow \nu_{s}$ oscillations, where at high
energies the matter effect suppresses oscillations.  Partially
contained events in Super-Kamiokande are estimated to be 97\% pure
$\nu_\mu$ charged current, with a mean neutrino energy of 10 GeV.  In
order to select higher energy $\nu_\mu$ events, which are more
sensitive to matter effects, we additionally require visible energy
greater than 5~GeV. We estimate the typical energy of the parent
atmospheric neutrino is 20~GeV.  After cuts are made upon the current
data sample we find 267 events.  Figure \ref{fig:MR}(c) shows the
zenith angle distribution of these events with predictions from MC, as
before.  Again we employ an up/down ratio to cancels systematic
uncertainties, with the same angular definition as used for the
multi-ring sample. There are 43 ``upward'' events and 84 ``downward''
events.  Figure \ref{fig:MR}(d) shows the $\Delta m^{2}$ dependence of
expected up/down ratio in the case of full mixing. For $\Delta m^2$ of
$3.2 \times 10^{-3}$~eV$^2$, the data are consistent with $\nu_{\mu}
\leftrightarrow \nu_{\tau}$ oscillation, whereas it differs from
$\nu_{\mu} \leftrightarrow \nu_{s}$ oscillation by 2.3 standard
deviations.


We estimated the total systematic uncertainty in the up/down ratio
to be $\pm4.1\%$, dominated by the $3.4\%$ uncertainty caused by the
mountain above the detector and the $2.0\%$ uncertainty caused by possible
background contamination by cosmic-ray muons. All other sources of
uncertainty were less than $1\%$.

\subsection{Upward Through-going Muon Sample}

Next we report on the search for possible matter effects by using upward
through-going muon events. The approach of this analysis is similar to that
for the PC events. Because the typical energy of the UTM parent neutrino is
approximately 100~GeV, matter effect suppression should appear most
prominently in this data set\footnote{Although stopping upward-going muons
  and PC events have similar parent neutrino energies, the limited zenith
  range of $-1 < \cos \Theta < 0$ with a relatively large uncertainty at the
  horizon\cite{SKupmu} limits their usefulness and they are not considered.}.
In the current sample, we have 1259.9 events after background subtraction.
Figure \ref{fig:MR}(e) shows the zenith angle distribution of these events
with predictions.  Again we utilize a ratio as the test parameter, dividing
``vertical'' and ``horizontal'' at cosine of zenith angle $= -0.4$. Figure
\ref{fig:MR}(f) shows the $\Delta m^{2}$ dependence of the expected
vertical/horizontal ratio in the case of full mixing. At the point of $3.2
\times 10^{-3}$~eV$^2$, the data are consistent with $\nu_{\mu}
\leftrightarrow \nu_{\tau}$ oscillation, while $\nu_{\mu} \leftrightarrow
\nu_{s}$ oscillation differs from the data by 2.9 standard deviations.


We estimated the total systematic uncertainty in the
horizontal/vertical flux ratio to be $\pm3.3\%$, dominated by the 3\%
uncertainty in the $\pi/K$ production ratio in the cosmic-ray
interaction in the atmosphere.  All other sources ofsystematic
uncertainty, including the background contamination in the most
horizontal bin\cite{SKupmu}, the spectral index of the neutrino flux,
and the difference between two independent flux
calculations\cite{Honda,Bartol} were 1\% or less.

\subsection{Combined Analysis}

Finally, we performed a combined statistical analysis of the
multi-ring, high energy partially contained, and upward-going muon
data sets.  For each sample ($i=$ MR,PC,UTM) we construct a one degree
of freedom $\chi^2_i$ defined by:
\begin{equation}
\chi^2_i = \frac{(N^A_{data}-\alpha_iN^A_{MC}(1+\frac{\epsilon_i}{2}))^2}
{(\sigma^A_{stat})^2} +  
\frac{(N^B_{data}-\alpha_iN^B_{MC}(1-\frac{\epsilon_i}{2}))^2}
{(\sigma^B_{stat})^2} + \frac{\epsilon^2_i}{\sigma^2_{i,sys}}
\end{equation}
\noindent where A is respectively either up-going or vertical and B is
down-going or horizontal. For the UTM sample, the calculated flux is
used in place of the number of events. The parameter denoting
normalization is $\alpha_i$, and for each ratio we introduce a
systematic uncertainty parameter $\epsilon_i$, weighted by the
estimated size of the uncertainty, $\sigma_{i,sys}$.

\begin{figure}[t]
\begin{center}
\psfig{figure=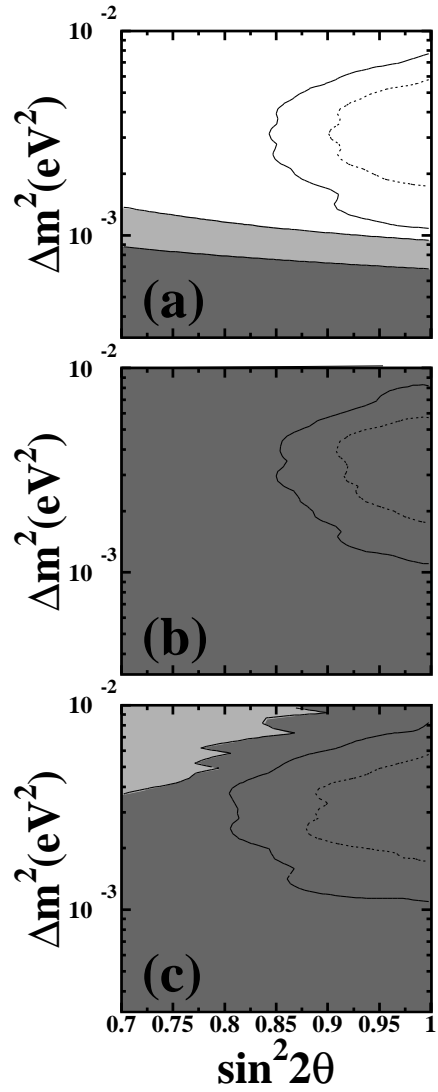,width=6.5in}
\caption{Excluded regions for three oscillation modes.
(a) $\nu_\mu \leftrightarrow \nu_\tau$, 
$\nu_\mu \leftrightarrow \nu_s$ with (b) \dmsq $>$ 0
and (c) \dmsq $<$ 0,
the light(dark) gray region
is excluded at 90(99)\% C.L., Thin dotted(solid) line indicates the
90(99)\% C.L. allowed regions from the analysis of FC single-ring
events.  }
\label{fig:combi} 
\end{center}
\end{figure}

The three $\chi^{2}_{i}$ are summed to form a total $\chi^{2}_{tot}$
with three degrees of freedom. The value of $\chi^{2}_{tot}$ is used
as a hypothesis test for the cases of $\nu_\mu \leftrightarrow
\nu_\tau$ and $\nu_\mu \leftrightarrow \nu_s$ oscillation. We exclude
regions in the $\sin^2 2\theta-\Delta m^2$ plane at the 90(99)\%
C.L. if the value of $\chi^2$ is greater than 6.3(11.3) for $\chi^2$
of three degrees of freedom.  Figure \ref{fig:combi} shows separately
the excluded regions for these three alternative oscillation modes,
along with the allowed region from FC single ring event analysis
\footnote{The two cases of $\Delta m^2 > 0$ and $\Delta m^2 < 0$ are treated
  continuously\cite{darkside}, using the minimum $\chi^2$ for the $\nu_{s}$
  fit which was slightly lower for the case of $\Delta m^2 > 0$, and this was
  used to draw the contours for both cases.}.  One sees that the parameters
allowed by the FC data in the $\sin^2 2\theta-\Delta m^2$ plane are excluded
at the 99\% confidence level by the independent tests for both positive and
negative $\Delta m^2$ sterile neutrino oscillations.

\section{Summary and Conclusion}


In summary, we have presented three independent data samples that
discriminate between the oscillations to either tau neutrinos or
sterile neutrinos in the region of mixing angle and $\Delta m^2$
preferred by the majority of the Super-Kamiokande data.  Two-flavor
oscillation between muon neutrinos and sterile neutrinos fit the low
energy charged current data, but do not fit the neutral current or
high energy data. We cannot exclude more complicated scenarios in
which both $\nu_{\mu} \leftrightarrow \nu_{\tau}$ and $\nu_{\mu}
\leftrightarrow \nu_{s}$ oscillations co-exist with small mixing to
sterile neutrinos, or with much smaller mass difference for sterile
neutrinos; yet there is nothing in this data to encourage one about
the existence of sterile neutrinos.  Pure $\nu_{\mu} \leftrightarrow
\nu_{\tau}$ neutrino oscillations fit all of the data samples
presented, without any inconsistency.

\section*{Acknowledgments}

We gratefully acknowledge the cooperation of the Kamioka Mining and Smelting
Company. The \SuperK{} experiment was built and has been operated with
funding from the Japanese Ministry of Education, Science, Sports and Culture,
and the United States Department of Energy. We gratefully acknowledge
individual support by the National Science Foundation, and the Polish
Committee for Scientific Research.

\newpage
\newpage


\end{document}